\documentclass[a4paper,twocolumn,accepted=2023-11-27]{quantumarticle}
\pdfoutput=1
\usepackage{graphicx}
\usepackage{layout}
\usepackage{dcolumn}
\usepackage{braket}
\usepackage{amsmath}
\usepackage{bm}
\usepackage{color}
\usepackage{verbatim}
\usepackage{hyperref}
\usepackage[utf8]{inputenc}
\usepackage[english]{babel}
\usepackage[T1]{fontenc}

\usepackage{tabularx}

\begin{document}

\title{The quartic Blochnium: an anharmonic quasicharge superconducting qubit}

\author{Luca Chirolli}
\affiliation{NEST Istituto Nanoscienze-CNR and Scuola Normale Superiore, I-56127 Pisa,
Italy}

\author{Matteo Carrega}
\affiliation{CNR-Spin, Via Dodecaneso 33, I-16146 Genova}

\author{Francesco Giazotto}
\affiliation{NEST Istituto Nanoscienze-CNR and Scuola Normale Superiore, I-56127 Pisa,
Italy}

\begin{abstract}
The quasicharge superconducting qubit realizes the dual of the transmon and shows strong robustness to flux and charge fluctuations thanks to a very large inductance closed on a Josephson junction. At the same time, a weak anharmonicity of the spectrum is inherited from the parent transmon, that introduces leakage errors and is prone to frequency crowding in multi-qubit setups. We propose a novel design that employs a quartic superinductor and confers a good degree of anharmonicity to the spectrum. The quartic regime is achieved through a properly designed chain of Josephson junction loops that shows minimal quantum fluctuations without introducing a severe dependence on the external fluxes. 
\end{abstract}

\maketitle

\section{Introduction} 

Superconducting qubits are nowadays one of the leading hardware platforms for quantum information and quantum computation purposes \cite{kjaergaard2020superconducting,huang2020superconducting,siddiqi2021engineering,bravyi2022future}.  In a future universal quantum computer, errors that originate from the interactions with the environment will be ideally completely eliminated via implementation of quantum error correction schemes \cite{gottesman1997stabilizer}. In the present context of the so-called noise intermediate-scale quantum (NISQ) era, it is very important to mitigate errors at the hardware level and quantum design represents a major resource in this sense. Since the construction of the prototypical charge box \cite{nakamura1999coherent,gambetta2017building}, the most successful superconducting qubit has been with no doubt the transmon \cite{koch2007charge-insensitive}, thanks to its intrinsic simplicity and its weak sensitivity to charge fluctuations, owing to a large shunt capacitance. Remarkable performances have also been demonstrated by the fluxonium \cite{manucharyan2009fluxonium,lin2018demonstration,nguyenPRX2019high-coherence,somoroff2021millisecond,zhang2021universal,nguyen2022blueprint}, that constitutes a highly coherent evolution of the flux qubit \cite{orlando1999superconducting} showing reduced sensitivity to charge noise thanks to the employment of a superinductance \cite{manucharyan2012thesis}, and represents a true competitor of the transmon. Multiple different designs of superconducting qubits have also been proposed, that aim at encoding an intrinsic protection in the wave function, and promising platforms definitely stand out, such as continuous variable qubits \cite{lloyd1999quantum,gottesman2001encoding,ofek2016extending,hu2019quantum,campagne-Ibarcq2020quantum}, $0-\pi$ and parity-protected qubits \cite{blatter2001design,protopopov2004anomalous,kitaev2006protected,gladchenko2009superconducting,smith2020superconducting,gyenis2021experimental,chirolli2021enhanced,chirolli2022swap,smith2022magnifying,calzona2023multi-mode,calzona2023anomalous}, and the bifluxon \cite{kalashnikov2020bifluxon}.

In the family of the  post-transmon setups employing superinductors \cite{bell2012quantum,bell2018josephson,peruzzo2020surpassing,feng-ming2021quantum,peruzzo2021geometric,ranadive2022kerr},  a particular instance is represented by the quasicharge superconducting qubit \cite{koch2009charging,pechenezhskiy2020superconducting}. Known also as Blochnium, it realizes the dual version of the transmon and achieves strong insensitivity to flux fluctuations thanks to a very large inductance, that plays the analogous role of a large shunt capacitance in the transmon. In addition, the presence of the superinductance renders the quasicharge degree of freedom dynamical, making the Blochnium also insensitive to charge fluctuations. This fundamental and intrinsic robustness comes at the expenses of a weak anharmonicity \cite{herrera-marti2013tradeoff}, that is directly inherited from the duality with the transmon. This aspect may become problematic due to frequency crowding in chips hosting several qubits, an issue that will become more and more important as the number of qubit on chip will increase, suggesting that novel approaches are necessary to improve the qubit performances. 

In this work we suggest to employ a quartic superinductor to strongly increase the anharmonicity of the Blochnium spectrum, thus complementing a circuit that is already insensitive to charge noise and very well protected from flux noise. A quartic inductor has been first proposed in Ref.~\cite{zorin2009superconducting} and recently realized in the quarton \cite{yan2020engineering} and the unimon \cite{hyyppa2022unimon} qubits. In all cases, the working principle is similar: a linear inductor in loop with a Josephson junction with a flux $\Phi_0/2$ threading the loop, with $\Phi_0=h/2e$, $h$ the Planck constant and $e$ the electron charge, cancels the quadratic term by destructive interference when the inductive energy matches the Josephson energy. In the unimon qubit this cancelation is achieved through an inductor long enough to match the inverse Josephson energy in the desired energy range, whereas in the quarton the inductor is formed by a chain of $M$ Josephson junctions. In this case, the $M$ junctions of the chain must have an energy $M$ times larger than the one they are in loop with. 

In our proposal, the quartic superinducting regime is achieved by designing a chain of quartons, that generalizes tunable non-linear superinductor designs previously proposed in the literature \cite{bell2012quantum,bell2018josephson,peruzzo2020surpassing,ranadive2022kerr}. The latter achieve the quartic superinducting regime with long chains of flux qubits \cite{orlando1999superconducting}, that represent the $M=2$ case. By employing a chain of $N$ quarton loops in series in the classical regime we achieve the potential 
\begin{equation}\label{Eq:Uphi}
U(\varphi)=\frac{E_{Jc}}{24MN^3}\varphi^4,
\end{equation}
that provides the system with an intrinsic nonlinearity, that awards a high degree of anharmonicity to the Blochnium spectrum. The quartic oscillator also induces a much larger spread of the wavefunction, with $\langle\varphi^2\rangle\gg 1$, that can be relevant to proposals for quantum computation based on grid states \cite{gottesman2001encoding, matveev2002persistent, brooks2013protected, smith2022magnifying}. Interestingly, since the Blochnium realizes the dual of the transmon, we effectively realize a quantum system described by a quartic kinetic term, something that is not very often encountered in physics.

As anticipated in Ref.~\cite{bell2018josephson}, in a general quartic superinductor quantum fluctuations induce a finite, quantum-limited linear inductance in the system, that represents the long wavelength mode of the system. We indeed find an inductance scaling $L\propto M^{1/3}N/\xi_J^{4/3}$, showing how quantum fluctuations can represent a severe limitation to a purely quartic regime and need to be properly accounted for.

The work is structured as follows: in Sec.~\ref{sect:4inductor} we introduce the model circuit and describe how to get a quartic superinductor out of it. Sec.~\ref{sect:4bloch} focuses on the properties of a Blochnium circuit built out of a quartic superinductor, while the impact of fluctuations in the electromagnetic environment and imperfections on the quartic blochnium are estimated in Sec.~\ref{sect:deco}. Sec.~\ref{sect:concl} contains our main conclusions and perspectives. Some technical details can be found in appendix.

\section{Quartic superinductor}
\label{sect:4inductor}

\begin{figure}[t!]
	\centering
	\includegraphics[width=1.0\linewidth]{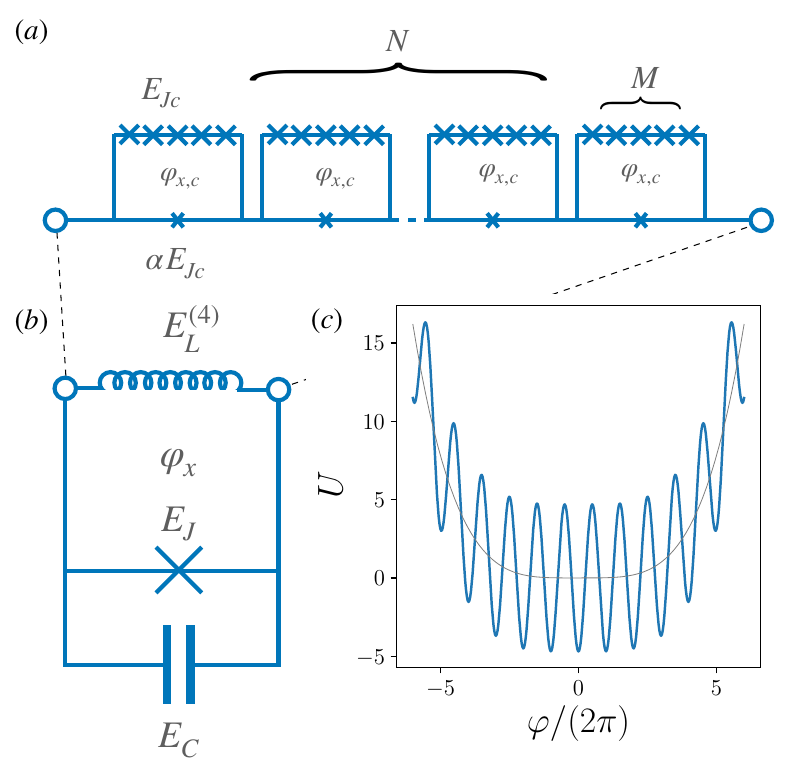}
	\caption{(a) Quartic superinductor constituted by a chain of $N$ unit cells, each one constituted by a loop with $M$ Josephson junctions of energy $E_{Jc}$ closed on a single Josephson junction with energy $\alpha E_{Jc}$, with $\alpha=1/M$. Each loop is threaded by a flux $\varphi_{x,c}=2\pi \Phi_{x,c}/\Phi_0=\pi$. (b) Quartic Blochnium circuits constituted by a quartic superinductor, realized as in (a) and described by a quartic inductive energy $E^{(4)}_L$, closed on a Josephson junction characterized by Josephson energy $E_J$ and charging energy $E_C$. (c) Effective Josephson potential of the quartic Blochnium experienced by the phase-particle in the circuit, for the choice $E_{J}=4.6~{\rm GHz}$, $E_{Jc}=100~{\rm GHz}$, $M=8$, and $N=40$ (in thin gray it is shown only the quartic dependence of the potential).}
	\label{Fig1}
\end{figure}

A superinductor is an inductive circuital element showing perfect DC conduction, extremely low dissipation, and a very high impedance that is able to exceed the resistance quantum at the relevant (microwave) frequency. Superinductors have been recently experimentally realized through long chains of nominally equal Josephson junctions \cite{manucharyan2012thesis,masluk2012microwave,bell2012quantum,niepce2019high,kuzmin2019quantum}, geometrical inductances \cite{peruzzo2020surpassing}, and granular materials \cite{grunhaupt2019granular,kamenov2020granular}.

In order to obtain a quartic superinductor we consider the system of Josephson junction depicted in Fig.~\ref{Fig1}(a). It is constituted by a chain of $N$ units, each one constituted by a loop consisting of $M+1$ Josephson junction, $M$ of which with equal Josephson energy $E_{Jc}$, and the last junction with energy $\alpha E_{Jc}$. Let us first consider the single unit cell. Denoting $\varphi'$ the phase difference across the small junction and $\varphi_n$ the phase difference across each junction, flux quantization imposes that $\varphi'+\sum_{n=1}^M\varphi_n+\varphi_{x,c}=2\pi n$, where $\varphi_{x,c}=2\pi \Phi_x/\Phi_0$ is the flux threading the loop in units of $\Phi_0/2\pi$. Introducing the total phase difference across the chain as $\varphi=\sum_{n=1}^M\varphi_n$, the full Josephson potential of the unit cell reads
\begin{eqnarray}\label{Eq:Uunitcell}
U_{\rm cell}&=&-E_{Jc}\sum_{n=1}^{M-1}\cos(\varphi_n)-E_{Jc}\cos\left(\varphi-\sum_{n=1}^{M-1}\varphi_n\right)\nonumber\\
&-&\alpha E_{Jc}\cos(\varphi+\varphi_{x,c}).
\end{eqnarray}
Choosing $\varphi_{x,c}=\pi$ and by  minimizing the potential for each phase $\varphi_n$ we find the superinductor solution $\varphi_n=\varphi/M$, and expanding the potential for small $\varphi$ the unit cell effective potential reads
\begin{equation}
\frac{U_{\rm cell}}{E_{Jc}}=\frac{\varphi^2}{2}\left(\frac{1}{M}-\alpha\right)+\frac{\varphi^4}{24}\left(\alpha-\frac{1}{M^3}\right).
\end{equation}
Clearly, for $\alpha=1/M$ the quadratic term is switched off and we are left with a quartic potential. This is what is used in the quarton \cite{yan2020engineering} as a qubit, together with a capacitive term that is associated to the effective junction. Assuming the latter to be negligible, we now consider a generic chain of $N$ loops. Denoting the overall phase drop across the chain as $\varphi$, the latter is divided to a good approximation in equal parts $\varphi/N$ across each loop and the effective superinductor potential takes the form 
\begin{equation}
U(\varphi)=\frac{E_{Jc}}{2N}\left(\frac{1}{M}-\alpha\right)\varphi^2+\frac{E_{Jc}}{24N^3}\left(\alpha-\frac{1}{M^3}\right)\varphi^4.
\end{equation}
For $\alpha=1/M$ we obtain the potential Eq.~\eqref{Eq:Uphi}. The great advantage of this procedure is that we can achieve a large quartic inductance $L_{(4)}\simeq MN^3$ without having $N,M$ to be very large. In fact, this means that with $N,M\simeq 10$ we already enter the Blochnium regime. This is particularly important because $\alpha$ is the ratio between the junction areas and the values of $M$ cannot be increased too much due to fabrication constraints.

\subsection{Quantum quartic oscillator}

For the choice $\alpha=1/M$ the quadratic term is suppressed and we end up with a purely quartic potential for the phase $\varphi$. Upon shunting the chain on a capacitor $C_0$ and quantizing the circuit by imposing canonical commutation rules between phase $\varphi$ and charge $q$, so that $q=-2ei\partial_\varphi$, we obtain a model of a quantum quartic oscillator
\begin{equation}\label{Eq:QuarticH}
H=-4E_C\frac{\partial^2}{\partial\varphi^2}+E^{(4)}_L\varphi^4,
\end{equation}
with $E_C=e^2/(2C_0)$ and $E_L^{(4)}=E_{Jc}/(24MN^3)$. The quartic oscillator has one very interesting property, in that already in absence of the periodic potential we have squeezed oscillator states.  The exact ground state of the Hamiltonian Eq.~\eqref{Eq:QuarticH} is not known analytically (see Appendix~\ref{App:QuarticO} for an expression of the spectrum), but we can use harmonic oscillator states as ansatz variational wavefunctions to estimate the spread of the wave function, $\xi\equiv \sqrt{\langle\varphi^2\rangle}$, and the spectrum. We introduce bosonic operators such that $\varphi=\frac{\xi}{\sqrt 2}(a+a^\dag)$ and $\partial /\partial \varphi=\frac{1}{\xi\sqrt{2}}(a-a^\dag)$. The value of $\xi$ is obtained by minimizing the expectation value of the Hamiltonian on the ground state $\psi_\xi(\varphi)=(\xi\sqrt{\pi})^{-1/2}e^{-\varphi^2/2\xi^2}$. For a harmonic potential ${\cal U}=E_L^{(2)}\varphi^2$, with $E_L^{(2)}=E_{Jc}/2N$, the ground state is exact and the spread of the wave function is $\xi_0=(4E_C/E_L^{(2)})^{1/4}=(8E_CN/E_{Jc})^{1/4}$, so that for a superinductor it grows as $N^{1/4}$. For the quartic oscillator, through the variational wave function approach we obtain $\xi=(4E_C/3E_L^{(4)})^{1/6}$, such that the spread of the wave function scales as
\begin{eqnarray}\label{Eq:widthXi}
\xi=\left(\frac{32ME_C}{E_{Jc}}\right)^{1/6}N^{1/2}.
\end{eqnarray}
This way, we are able to change the behavior of the phase fluctuations from $\langle\varphi^2\rangle\simeq (N)^{1/2}$ to $\langle\varphi^2\rangle\simeq N$.

\section{Quartic Blochnium}\label{sect:4bloch}

\begin{figure*}
	\centering
	\includegraphics[width=1.\textwidth]{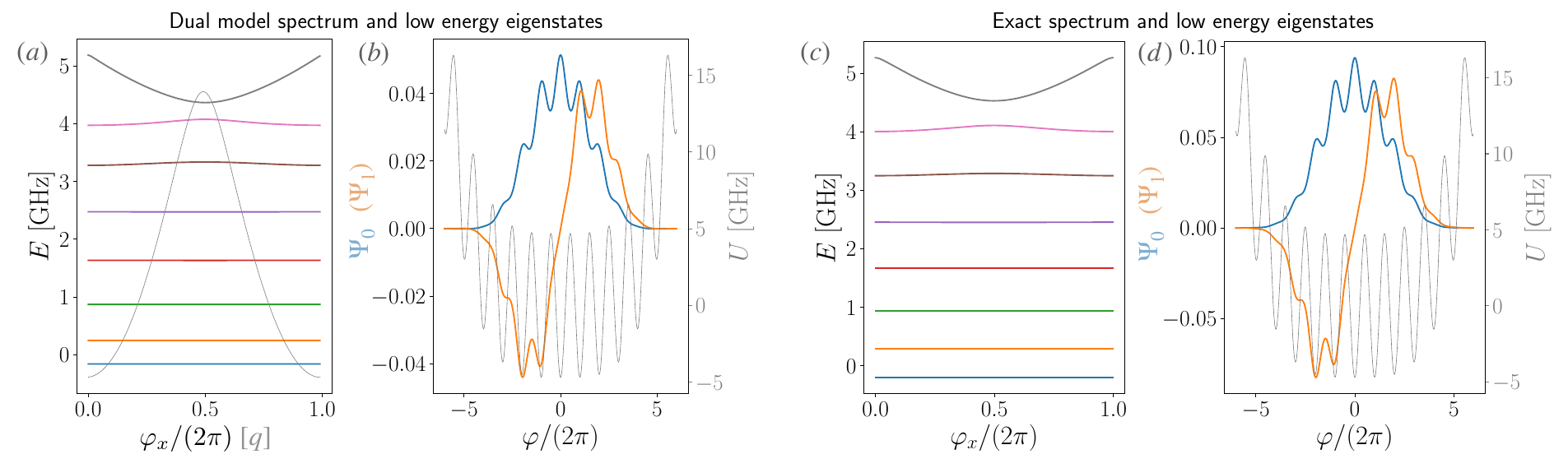}
	\caption{(a) Blochnium low energy spectrum versus external flux obtained by diagonalization of the dual model Eq.~(\ref{Eq:DualHam}). In light gray the first quasicharge band dispersion $\epsilon_0(q)$ of the transmon-like problem. The parameters of the model are $E_{J}=4.6~{\rm GHz}$, $E_{Jc}=100~{\rm GHz}$, $E_C=7.0~{\rm GHz}$, $M=8$, and $N=40$, for which $E_L^{(4)}\simeq 8\times 10^{-6}~{\rm GHz}$. (b) Ground state and first excited state  wave functions (left axis) and Josephson potential (thin gray, right axis). (c) and (d) same as (a) and (b), respectively, for the exact Hamiltonian Eq.~(\ref{HamiltoaniQBphi}).}
	\label{Fig2}
\end{figure*}

We now consider the circuit formed by closing the chain on a Josephson junction characterized by capacitance $C_0$ and Josephson energy $E_J$, as shown  in Fig.~\ref{Fig1}b). Setting $\alpha=1/M$ we switch the quadratic term off and we are left with the quartic potential $U=E_L^{(4)}\varphi^4$. Again, the Hamiltonian is then quantized by imposing canonical commutation rules between phase $\varphi$ and charge $q$, so that $q=-2ei\partial_\varphi$, and the quartic Blochnium Hamiltonian reads
\begin{equation}\label{HamiltoaniQBphi}
H_B=-4E_C\frac{\partial^2}{\partial\varphi^2}-E_J\cos(\varphi_x-\varphi)+E_L^{(4)}\varphi^4,
\end{equation}
where $E_C=e^2/(2C_B)$ is the effective charging energy in terms of the effective capacitance $C_B$ associated to the additional junction. The full Josephson potential is shown in Fig.~\ref{Fig1}(c) for an effective inductive energy $E^{(4)}_{Jc}=8\times 10^{-6}~{\rm GHz}$, obtained for a choice of parameter $E_{Jc}=100~{\rm GHz}$, $M=8$, and $N=40$, and $E_J=4.7~{\rm GHz}$. 

The Blochnium regime is achieved when $E^{(4)}_L\ll E_C,E_J$ and $E_C\sim E_J$, in which case by neglecting the inductive term we obtain solutions of the Schr\"odinger equation  in the form of Bloch waves, that represent quasicharge states. We then consider the action of the inductive term in generating a long wavelength modulation of the weight of these quasicharge states, in an envelope function approximation. The latter is achieved by writing the wave function as
\begin{equation}\label{Phi-varphi}
\Psi_{l,0}(\varphi)=\int_{-1/2}^{1/2}dq v_{l,0}(q)\psi_{q0}(\varphi),
\end{equation}
where $\psi_{q,0}(\varphi)=e^{iq\varphi}u_{0q}(\varphi)$ is a Bloch function satisfying $(-4E_C\partial^2_\varphi-E_J\cos(\varphi))\psi_{q,0}(\varphi)=\epsilon_0(q)\psi_{q,0}(\varphi)$. For the lowest band we can approximate $\epsilon_0(q)=-\lambda_0\cos(2\pi q)/2$, with $\lambda_0=\frac{8}{\sqrt{\pi}}(8E_J^3E_C)^{1/4}e^{-\sqrt{8E_J/E_C}}$ the bandwidth of the transmon charge dispersion. Neglecting the coupling between the $s=0$ band and the $s>0$ \cite{koch2009charging} bands, the weight $v_{l,0}(q)$ satisfies the equation
\begin{equation}\label{Eq:DualHam}
\left[E^{(4)}_L\left(i\frac{d}{dq}-\varphi_x\right)^4-\frac{\lambda_0}{2}\cos(2\pi q)-E_{l,0}\right]v_{l,0}(q)=0.
\end{equation}
We clearly see that the effective Hamiltonian for the weights $v_{l,0}(q)$ is dual to the transmon Hamiltonian and features a quartic derivative rather than a second one. 

An analytic approximation of the spectrum can be obtained in the spirit of the quantum quartic oscillator previously outlined. We expand the cosine potential at small momentum and assume harmonic oscillator wave functions with a spread in quasicharge $\sigma$ to be determined by energy minimization. If the spread $\xi$ in phase is enhanced by the quartic potential, the spread $\sigma$ in quasicharge is expected to shrink, and indeed we find
\begin{equation}\label{Eq:SigmaVsN}
\sigma=\left(\frac{3E_L^{(4)}}{2\pi^2\lambda_0}\right)^{1/6}=\frac{1}{\sqrt{N}}\left(\frac{E_{Jc}}{16\pi^2\lambda_0M}\right)^{1/6}.
\end{equation}
We can then use the value of $\sigma$ to estimate the energy difference between the ground state and the first excited state, obtaining 
\begin{equation}
E_{01}=\lambda_0\left(\frac{96\pi^4 E^{(4)}_L}{\lambda_0}\right)^{1/3}=\frac{\lambda_0}{N}\left(\frac{4\pi^4 E_{Jc}}{\lambda_0 M}\right)^{1/3}.
\end{equation}
Assuming $E_C\simeq E_J$, that is the relevant working regime for the Blochnium,  we have $\lambda_0\simeq E_C,E_J$ (reasonable values are $E_C,E_J\simeq 1\div 10 ~ {\rm GHz}$). The relevant working regime to have a transition frequency in the GHz spectrum is provided by the choice $E_{Jc}\simeq 100 \lambda_0$, that is compatible with a choice of $N,M\simeq 10$, and $\alpha=1/M$.

In order to check the analytical predictions we numerically calculate the spectrum and the wave functions of the quartic Blochnium in the dual representation. We expand in Fourier series $u_0(q)=\sum_n e^{inq}u_{0,n}$ and obtain the dispersion $\epsilon_0(q)$ through numerical diagonalization of the transmon Hamiltonian with $E_C=7.0~{\rm GHz}$ and $E_J=4.7~{\rm GHz}$. The resulting first quasicharge band is shown in Fig.~\ref{Fig2}a), thin gray curve, and is well approximated by a two-harmonic expression $\epsilon_0(q)=-\lambda_0\cos(2\pi q)/2+\lambda_1\cos(4\pi q)/2$. We then expand in Fourier series $v_{l,0}(q)=\sum_m e^{-2\pi imq}v_{l,m}$ and numerically diagonalize the Hamiltonian Eq.~\eqref{Eq:DualHam} with the quartic potential depicted in Fig.~\ref{Fig2}(b) (thin gray, right axis) has a quartic derivative (care has to be paid to a possibly $q$-dependent phase of $\psi_{q,0}(\varphi)$). The resulting spectrum is shown in Fig.~\ref{Fig2}a): we clearly see an evident spectrum anharmonicity, as expected from the quartic character of the potential. Furthermore, given the parameter choice, the lowest states show a very flat dependence on the flux $\varphi_x$ threading the Blochnium circuit, in agreement with the expectation of a flux insensitive regime (without considering variations of $\varphi_{x,c}$). The higher excited states show reduced flatness and a weak sensitivity to the external flux. The ground state and first excited state wave functions are shown in Fig.~\ref{Fig2}(b). 

In order to check the accuracy of the dual model Eq.~\eqref{Eq:DualHam}, we numerically diagonalize the exact Hamiltonian Eq.~\eqref{HamiltoaniQBphi} and report the spectrum for comparison in Fig.~\ref{Fig2}(c). The similarity is very good, confirming the validity of the dual representation for the parameter regime. The exact ground state and first excited state wave functions are shown in Fig.~\ref{Fig2}(d), that also confirm the validity of the model.

An estimate of the expected spread of the wave function in $\varphi$ space gives $\xi/(2\pi)\simeq 1.6$, that is compatible with an eye inspection. An important point of the Blochnium physics, and that somewhat defines its working regime, is that the wave function spreads among several wells of the cosine potential. It is this property that confers robustness to flux noise.  Interestingly, we notice that in the original Blochnium circuit employing a linear inductance $E^{(2)}_L\propto 1/N$, assuming $E_C\propto 1/N$ due to stray capacitances yields a spread $\xi_0$ that is $N$-independent and fixed by the sample parameters. In contrast, Eq.~\eqref{Eq:widthXi} shows that in our setup the effective spread $\xi_{\rm eff}\propto N^{1/3}$, showing that it is possible to enter the Blochnium regime by increasing $N$, regardless of the possible presence of stray capacitances.

\subsection{Anharmonicity}

\begin{figure}
	\centering
	\includegraphics[width=1.0\linewidth]{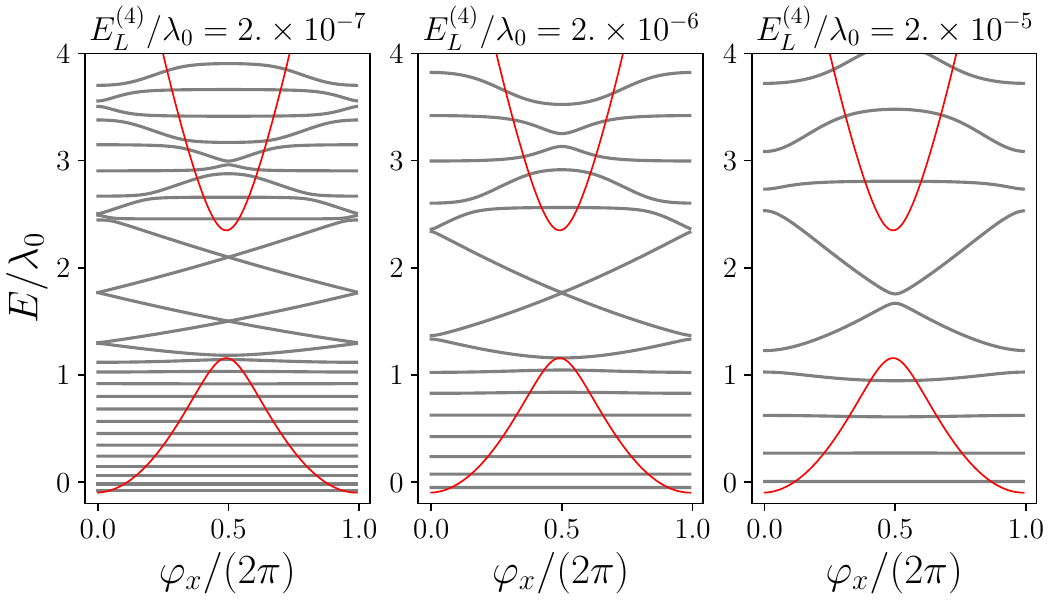}
	\caption{Spectrum of the quartic Blochnium as a function of the external flux $\varphi_x$, for three different values of $E_L^{(4)}$ and by keeping constant $E_J=4.0~{\rm GHz}$ and $E_C=7.0~{\rm GHz}$, from  which it follows that $\lambda_0=3.94~{\rm GHz}$. Superimposed are the first two Bloch bands of the parent transmon problem.}
	\label{Fig:Spectrum}
\end{figure}

It is instructive to study the spectrum of the Hamiltonian Eq.~\eqref{HamiltoaniQBphi} as a function of $E_L^{(4)}$, to check the validity of the Blochnium regime. In Fig.~\ref{Fig:Spectrum} we present the spectrum as a function of the external flux $\varphi_x$ for three values of $E_L^{(4)}$. We clearly see that the Blochnium regime, characterized by quasi-flat states in the allowed energies of the transmon bands and dispersive states in the gaps between the transmon bands, is well achieved for small values $E_L^{(4)}/\lambda_0$. In turn, by increasing the ratio $E_L^{(4)}/\lambda_0$ the Blochnium regime is gradually lost and all states become flux dependent.

\begin{figure}[t]
	\centering
	\includegraphics[width=1.0\linewidth]{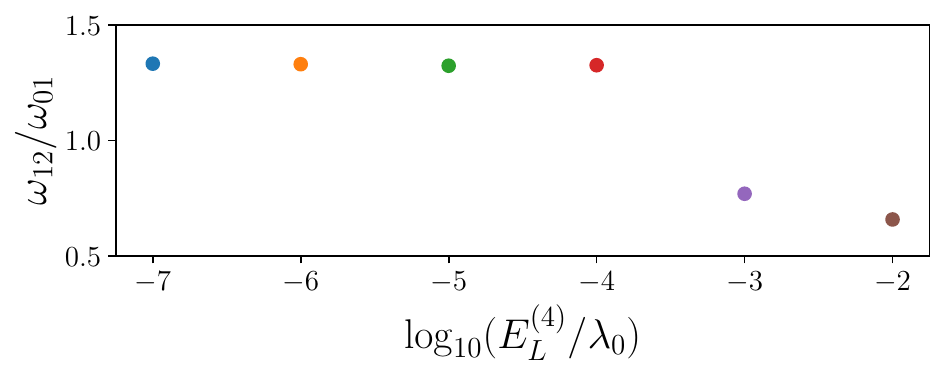}
	\caption{Anharmonicity of the quartic Blochnium, $\omega_{12}/\omega_{01}$, as a function of the ratio $E_L^{(4)}/\lambda_0$, showing the intrinsic constant anharmonicity character for values $E_L^{(4)}/\lambda_0< 10^{-4}$.}
	\label{Fig:anharmonicity}
\end{figure}

It is customary to quantify the anharmonicity of a qubit by comparing the transition between the first and second excited states, $\omega_{12}$, with the transition between the ground and first excited state, $\omega_{01}$. In the quartic potential approximation Eq.~\eqref{HamiltoaniQBphi}, the anharmonicity of the spectrum is intrinsic. An estimate can be performed at the level of the dual Hamiltonian Eq.~\eqref{Eq:DualHam} by expanding the cosine potential for $E_L^{(4)}/\lambda_0\ll 1$ and by employing the spectrum of the quartic oscillator provided in Appendix \ref{App:QuarticO}, that gives $E_n/\lambda_0\simeq (3\pi^4E_L^{(4)}/2\lambda_0)^{1/3}(3/2+10n/3+n^2)$. The result is an intrinsic anharmonicity 
\begin{equation}
\frac{\omega_{12}}{\omega_{01}}=1+6/13,
\end{equation}
giving about 50\% qubit anharmonicity. As shown in Fig.~\ref{Fig:anharmonicity}, this rough estimate agrees very well with numerical calculation of the anharmonicity through full diagonalization of Eq.~\eqref{HamiltoaniQBphi} up to a value $E_L^{(4)}/\lambda_0<10^{-4}$, beyond which the Blochnium regime is lost. 

\begin{figure}[t]
	\centering
	\includegraphics[width=1.0\linewidth]{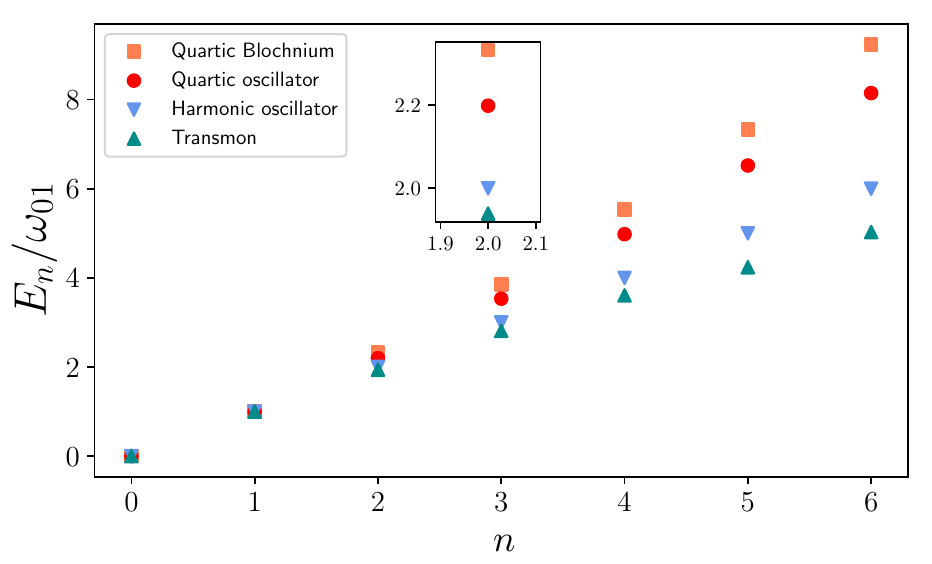}
	\caption{Spectrum of the quartic Blochnium, shown together with the spectrum of a transmon characterized by $E_J/E_C=50$, the spectrum of a quantum harmonic oscillator, and the spectrum of a quantum quartic oscillator, each normalized to its $\omega_{01}$ transition. Inset: zoom on the $\omega_{12}$ transition, showing the anharmonicity induced by the different terms.}
	\label{Fig3}
\end{figure}

In order to further emphasize the anharmonicity of the quartic Blochnium spectrum, in Fig.~\ref{Fig3} we compare it with the spectrum of a conventional transmon characterized by $E_J/E_C=50$, the spectrum of the quantum harmonic oscillator, and the spectrum of a quantum quartic oscillator, each normalized by its associated transition $\omega_{01}$. On the scale of six excited states we see that the transition $\omega_{12}/\omega_{01}$ (see inset) is clearly resolved in the case of the quartic Blochnium and the case of a quartic oscillator, whereas it is hardly resolved in the case of a transmon with respect to the harmonic oscillator case. This shows that a quartic anharmonicity is sufficient to obtain a good addressability of the $\omega_{01}$ transition and promotes quartic potentials as a relevant route to fight leakage errors in contrast to strongly anharmonic flux-dependent designs typical of the flux qubit and fluxonium design, that necessarily introduce strong sensitivity flux fluctuations.

\section{Collective modes in the superinductor}

We now assess the collective modes in the Josephson junction chain and their role in the dynamics of the Blochnium circuit. The collective modes will be studied by first discussing the behavior of the single unit cell, that realizes a single quarton circuit, and then by considering the entire structure composed by a chain of quartons. Finally an analysis of the impact of quantum fluctuations on the low energy degrees of freedom will be presented.

\subsection{Quarton unit cell}

The single unit cell of the chain constitutes a quarton qubit, that is formed by $M$ Josephson junctions in series closed on a junction of size reduced by a factor $\alpha$. The independent degrees of freedom are constituted by the phase differences $\varphi_n$ at the $M$ junctions and the dynamics of the circuits is best spelt in terms of collective modes.  The circuit is characterized by mainly three capacitances, the capacitance of the $M$ equal junctions, $C_J$, the one of the smaller junction, $\alpha C_J$, and the one to ground of each superconducting island, $C_g$. Expanding the Josephson potential up to second order in the phases $\varphi_n$ we can study the spectrum of the plasma oscillations. 

In order to single out the low energy mode, following Ref.~\cite{fergusonPRX2013symmetries} we introduce a set of specific collective modes, the {\it average mode} $\varphi=\sum_{n=1}^{M}\varphi_n$ and the remaining $M-1$ {\it difference modes} $\gamma_l$, so that we can express the local phase differences as
\begin{equation}\label{Eq:Wmodes}
\varphi_n=\varphi/M+\sum_\mu W_{\mu n}\gamma_\mu,
\end{equation}
with $W_{\mu m}=\sqrt{2/M}\cos\left[\pi \mu(m-1/2)/M\right]$ for $m=1,\ldots, M$ and $\mu=1,\ldots,M-1$. The potential is then expanded at fourth order in $\varphi$ and second order in $\gamma_\mu$, in a way that the Hamiltonian for the low energy mode $\{\varphi,q\}$ can be written as
\begin{equation}
H_Q=\frac{q^2}{2C_Q}+E^{(2)}_{L_Q}\varphi^2+\frac{E_{Jc}}{24M}\varphi^4,
\end{equation}
with $E^{(2)}_{L_Q}=E_{Jc}(1/M-\alpha)/2$. For the choice $\alpha=1/M$ the quadratic term vanishes exactly and results in the quarton Hamiltonian \cite{yan2020engineering}.  

\begin{figure}[t]
	\centering
	\includegraphics[width=1.0\linewidth]{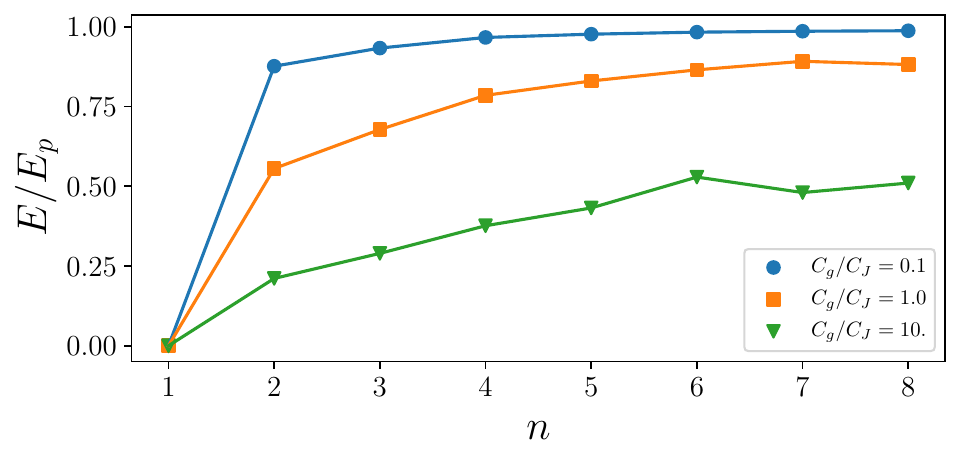}
	\caption{Quarton collective modes  in units of $E_p=\sqrt{8E_{C_J}E_{Jc}}$ for $M=8$, $\alpha=1/8$, and three values of $C_g/C_J$. In the quadratic 			approximation the first mode has exactly zero energy.}
	\label{Fig6}
\end{figure}

The other $M-1$ modes are described by a quadratic Hamiltonian accounting for the higher energy plasma modes. In Fig.~\ref{Fig6}, we report the results for $\alpha=1/M$ for different values of $C_g/C_J$ and for the choice $M=8$. The spectrum approximately follows the continuum expression \cite{masluk2012microwave,weiss2015kerr}, with the first mode pinned at zero energy for all values of $C_g/C_J$, as a result of the cancelation at flux $\Phi_0/2$. In addition, we see that for large values of $C_g/C_J$ all modes experience a reduction of their energy, whereas small values of the ground capacitance keep all but the first mode close to the plasma frequency $E_p=\sqrt{8E_{C_J}E_{Jc}}$, with $E_{C_J}=e^2/(2C_J)$. In order to separate the dynamics of low and high energy modes, we ask the high energy modes to be as close as possible to the plasma frequency and obtain a constraint for the maximum $C_g$, 
\begin{equation}
\left(\frac{C_g}{C_J}\right)^{1/2}<\frac{1}{M},
\end{equation}
in agreement with \cite{manucharyan2009fluxonium}.

Neglecting the capacitive coupling between the difference modes charges $p_\mu$ and the average charge $q$ in the $C_g/C_J\ll 1$ limit, higher and low energy modes are coupled by a residual term that originates from the quartic interaction and is well approximated by 
\begin{equation} 
H_c=-\frac{E_{Jc}}{4M^2}\varphi^2\sum_\mu \gamma^2_\mu.
\end{equation}
Assuming all difference modes to be not excited and by taking their expectation value on the Gaussian ground state (or thermal state), we immediately obtain a correction to the quadratic term of the low energy average mode $\varphi$, so that 
\begin{equation}
E^{(2)}_{L_Q}=\frac{E_{Jc}}{2}\left(\frac{1-\xi^2_J/4}{M}-\alpha\right),
\end{equation}
where $\xi^2_J=\sqrt{8E_{C_J}/E_{Jc}}$. The true quarton Hamiltonian \cite{yan2020engineering} is then obtained with a slightly smaller junction, characterized by $\alpha=(1-\xi^2_J/4)/M$, properly corrected by quantum fluctuations. 
This way, the unit cell Hamiltonian is well approximated by
\begin{eqnarray}\label{Eq:Hquarton}
H_{\rm cell}(\varphi,q)&=&\frac{q^2}{2C_Q}+\frac{E_{Jc}\varphi^4}{24M},
\end{eqnarray}
with $C_Q=2C_J/M$ for $C_g/C_J\ll 1$ and $C_Q=C_g/2$ for $C_J/C_g\ll 1$, and the other difference modes can be neglected.

\subsection{Quarton chain}

The very same analysis can be repeated for the chain of quartons that constitute the entire non-linear superinductor and is described by the Hamiltonian
\begin{equation}
H=\sum_{j=1}^NH_{\rm cell}(\varphi_j,q_j).
\end{equation}
By neglecting a capacitive coupling to ground of each island separating two quartons, the system is composed by a collection of uncoupled quartic oscillators and the general wave function will be given by a product state
\begin{equation}
\Psi_{\{l\}}(\boldsymbol{\varphi})=\prod_{j=1}^N\chi_{l_j}(\varphi_j),
\end{equation}
with the $\chi_l$ eigenfunctions of the  Hamiltonian Eq.~(\ref{Eq:Hquarton}). In the ground state, we can assume the quartic oscillators to be all in a Gaussian state, described by a spread $\xi_q$ that is determined by energy minimization, 
\begin{equation}
\xi^2_q=2M^{2/3}\xi_J^{4/3}.
\end{equation}

When closing the chain on the Blochnium small Josephson junction, the collective modes become the relevant degrees of freedom. We then introduce the {\it superinductance mode} $\varphi=\sum_{j=1}^N\varphi_j$ and difference modes $\gamma_\nu=\sum_jV_{\nu j}\varphi_j$ in analogy with Eq.~\eqref{Eq:Wmodes}. We can express the wave function in the new variables and since the difference modes are orthogonal to the superinductance mode, no mixing arises in the Gaussian state, that can be written as 
\begin{equation}
\Psi=e^{-\varphi^2/(2N\xi_q^2)}\prod_{\nu=1}^{N-1}e^{-\gamma_\nu^2/2\xi_q^2}.
\end{equation}
We clearly see that the spread of the wave function in the superinductance mode is rescaled as $\xi^2_q\to N\xi^2_q$, in agreement with the typical behavior of the wave function in case of a linear superinductance. Although formally exact, this step represents a strong approximation for the wave function of the chain of quartic oscillators. 

When closing the chain on the small Blochnium Josephson junction, the wave function keeps the factorizes form, with the Gaussian state for the difference mode $\gamma_\nu$ unchanged, and the Gaussian state of the superinductance mode $\varphi$  replaced by the Blochnium wave function
\begin{equation}
e^{-\varphi^2/(2N\xi_q^2)} \to \Psi_B(\varphi),
\end{equation}
with the Blochnium wave function localized on a length scale $\xi$ given in Eq.~\eqref{Eq:widthXi} and the Gaussian wave function localized on a much shorter length scale $\xi_q\ll \xi$. This point is crucial and represents the core of the superinductance classical approximation, that is valid in the case $E_{C_J}/E_{Jc}\to 0$ and neglects the contribution of quantum fluctuation.

\subsection{Quantum fluctuations}

For non-negligible $E_{C_J}$, we expect that the non-linear coupling between the $\gamma_\mu$ and $\varphi$ modes introduces a correction to the effective quadratic term, as in the case discussed for the quarton in the previous section. The Hamiltonian for the chain of quartons expressed in terms of the $\varphi$ and $\gamma_\nu$ modes reads
\begin{eqnarray}
H&=&\sum_{\nu=1}^{N-1}\left[\frac{p_\nu^2}{2C_Q}+\frac{E_{Jc}\gamma_\nu^4}{16MN}\right]+\frac{E_{Jc}}{8MN}\sum_{\nu\neq \nu'}\gamma_\nu^2\gamma_{\nu'}^2\nonumber\\
&+&\frac{Nq^2}{2C_Q}+\frac{E_{Jc}\varphi^4}{24MN^3}+\frac{E_{Jc}\varphi^2}{4MN^2}\sum_{\nu=1}^{N-1}\gamma_\nu^2,
\end{eqnarray}
and we have neglected odd parity coupling between the different modes. First of all, by neglecting the coupling between different modes we see that although the spectrum of the $\varphi$ and $\gamma_\nu$ modes is characterized by the same energy scale 
\begin{equation}
\Delta\sim \left(\frac{3M}{64N}\right)^{1/3}(E_{C_J}^2E_{Jc})^{1/3}, 
\end{equation}
the spread of the wave function in the superinductance and difference modes show a different scaling with $N$, 
\begin{equation}
\xi_\varphi^2=N\xi^2_q, \qquad \xi^2_\gamma=(N/6)^{1/3}\xi^2_q
\end{equation}
suggesting that a separation of scales does indeed occur. 

In order to assess the impact of the coupling between the modes, we can once again assume the ground state to be described by a Gaussian wave function for the modes $\gamma_\nu$ and calculate its spread by energy minimization. This way, we find
\begin{equation}
\xi^2_\gamma=\left(\frac{N}{3/2+N-2}\right)^{1/3}\xi_q^2.
\end{equation}
It follows that due to the repulsive interaction between the modes, the spread of the wave function $\xi_\gamma$ shrinks. By taking the expectation value of the Hamiltonian on the Gaussian wave function of the $\gamma_\nu$ modes we finally find
\begin{equation}
H_{\rm chain}=\frac{Nq^2}{2C_Q}+\frac{E_{Jc}\varphi^4}{24MN^3}+\frac{(3/2)^{1/3}E_{Jc}\xi_J^{4/3}}{2NM^{1/3}}\varphi^2,
\end{equation}
showing how a quadratic term is generated for the superinductance mode, whose origin is purely due to quantum fluctuations. We then conclude that the quadratic term renormalization can be suppressed by reducing $\xi_J$, that can be achieved by reducing the charging energy $E_{C_J}$ through proper shunting capacitances. 

At this point, we can make contact with previous and similar proposals of a non-linear superinductor realized in a similar way as a chain of flux qubits \cite{bell2012quantum,bell2018josephson}, that is a particular case of $M=2,3$ depending on the flux qubit realization. Analogously, we find a fluctuation induced inductive term that is due to higher energy modes that are coupled to the low energy one due to the non-linearities. The resulting fluctuation-induced inductance scales with the number of loops and junctions as $NM^{1/3}$, offering us an insight to assess its overall impact.

\section{Decoherence estimates}
\label{sect:deco}

The spectrum of the quartic Blochnium shows strong insensitivity of the lowest energy states to the flux $\varphi_x$ threading the main loop that, together with the intrinsic insensitivity to charge fluctuations, renders the qubit particularly appealing. Nevertheless, it is important to assess the impact of fluctuations in the electromagnetic environment and imperfections of the chain. The latter statically couple to $\varphi^2$ and their overall magnitude statistically averages to zero. Much more important are fluctuations in the electromagnetic environment in the form of time-dependent variations of the fluxes that thread the loops of the quartic superinductor and the offset charges localized on the superconducting islands separated by the Josephson junctions. 

In general, given the two lowest energy states of the qubit, $|0\rangle$ and $|1\rangle$, and given a classical randomly fluctuating variable $\lambda_i$ linearly coupled to the observable ${\cal O}_i$ through the Hamiltonian $H=\sum_i\lambda_i{\cal O}_i$, the relaxation rate at temperature lower than the qubit frequency can be estimated through Fermi golden rule \cite{chirolli2006decoherence}
\begin{equation}
\Gamma_1=\sum_i|\langle0|{\cal O}_i|1\rangle|^2S_i(\omega_{01}),
\end{equation}
where $S_i(\omega)$ is the spectral function of the fluctuating variable $\lambda_i$, $S_i(\omega)=\int dt e^{-i\omega t}\langle \lambda_i(0)\lambda_i(t)\rangle$. 

The dephasing rate depends on the characteristic time scales of the spectral function, in that it is affected by the zero-frequency part of it. For well behaved $S(\omega=0)$ the decay of the off-diagonal matrix element of the qubit density matrix is exponential, with a rate
\begin{equation}
\Gamma_\phi=2k_BT\sum_i\left|\langle1|{\cal O}_i|1\rangle-\langle0|{\cal O}_i|0\rangle\right|^2\left.\frac{S_i(\omega)}{\omega}\right|_{\omega\to 0}.
\end{equation}
In turn, for spectra singular at $\omega=0$ such as $1/f$ noise described by $S_{\lambda_i}(\omega)=2\pi A^2/|\omega|$, the decay is Gaussian over a time scale dictated by the inverse of  \cite{koch2007charge-insensitive} 
\begin{equation}\label{Eq:Gamma-phi}
\Gamma_\phi=A\sum_i\left|\frac{\partial E_{01}}{\partial \lambda_i}\right|,
\end{equation}
that collects all rates independently.

\subsection{Flux fluctuations}

Time-dependently fluctuating fluxes in the loops of the superinductor chain enter the Hamiltonian via the perturbation $\delta {\cal U}$,
\begin{eqnarray}\label{Eq:PhiX-sw}
\delta{\cal U}(t)=-\alpha E_{Jc}\frac{\varphi}{MN}\sum_{j=1}^{N}\delta\varphi_{x,j}(t).
\end{eqnarray}
The operator $\varphi$ affects the relaxation rate by inducing transitions between the two qubit states, but its odd character has zero expectation value on the qubit states, so it has no effect on the dephasing rate at lowest order. Fluctuations in the flux $\delta\varphi_{x,j}$ can arise either due to short wavelength fluctuations of the magnetic fields or due to long wavelength fluctuations in presence of random area variations among the loops, $\delta\varphi_{x,j}=2\pi(A\delta B_j+\delta A_j B)/\Phi_0$. Short wavelength fluctuations of the field are expected to be correlated on a short-range scale and add up incoherently, thus producing an overall factor $N$. The full relaxation rate is estimated as
\begin{equation}\label{Gamma1-flux}
\Gamma_1=\left(\frac{4\pi \alpha E_{Jc}}{MN\Phi_0}\right)^2|\langle0|\varphi|1\rangle|^2S_{\delta\Phi}(\omega_{01}),
\end{equation}
where $S_{\delta\Phi}(\omega)=\int dt e^{-i\omega t}\langle \delta\Phi_x(t)\delta\Phi_x(0)\rangle$ is the spectral function of the fluctuating flux $\delta\Phi_x=(\Phi_0/2\pi)\sum_j\delta\varphi_{x,j}$ and the relevant matrix element is given by (see Appendix \ref{App:MatrixEl})
\begin{equation}
\langle 0|\varphi|1\rangle=\frac{i}{\sqrt{2}\sigma},
\end{equation}
with $\sigma$ given in Eq.~\eqref{Eq:SigmaVsN}. The transition matrix element thus scales with the inverse of the quasicharge spread of the wave function and is proportional to $\sqrt{N}$. This can be understood as arising from the enhanced spread of the wave function, that yields a larger dipole matrix element. As a result its square cancels one power of $N$ in the rate Eq.~\eqref{Gamma1-flux}, that becomes
\begin{equation}
\Gamma_1=\frac{4\pi^2}{NM^4}(2\pi^2\lambda_0ME^5_{Jc})^{1/3}\frac{S_{\delta\Phi}(\omega_{01})}{\Phi^2_0}.
\end{equation}

For the case of the fluxonium, that is also composed by a chain of Josephson junctions, it is customary to study the decoherence rate as a function of the number of junctions by keeping constant the energy scales defining the problem \cite{manucharyan2009fluxonium}. In our case, by fixing $E_J$, $E_C$, $E_L^{(4)}$, and setting $\alpha=1/M$, the relaxation rate due to fluctuations in the loops becomes
\begin{equation}
\Gamma_1=c\frac{N^4}{M^2}\left(\lambda_0(E_L^{(4)})^5\right)^{1/3}\frac{S_{\delta\Phi}(\omega_{01})}{\Phi_0^2},
\end{equation}
where $c=1536(3\sqrt{2})^{2/3}\pi^{8/3}\simeq 0.85\times 10^5$, that predicts an increase of the relaxation rate with the fourth power of the number of loops. This result owes its origin to non-linear dependence of $E_L^{(4)}$ on $N$. Using values $E_L^{(4)}\simeq 10^{-6}~{\rm GHz}$ and $\lambda_0\simeq E_C,E_J\simeq {\rm GHz}$ we find that $\Gamma_1\simeq 10^{-5}(N^4/M^2)\times{\rm GHz}^2\times  S_{\delta \Phi_j}/\Phi_0^2$. 

As an estimate of the worst scenario case, we assume a $1/f$ noise power spectrum affecting global field fluctuations for which $S_\Phi=\sum_{ij}\langle \delta A_i\delta A_j/A^2\rangle A_\Phi^2/\hbar\omega$ with $A_\Phi=10^{-6}\Phi_0$ \cite{yan2016theflux,nguyenPRX2019high-coherence,nguyen2022blueprint}. Statistically averaging over an area error of 10\% gives a factor $\sum_{ij}\langle \delta A_i\delta A_j/A^2\rangle=N(0.1)^2$, and for a qubit frequency in the range of GHz we obtain $\Gamma_1\simeq N^5/M^2\times 10^{-10}{\rm Hz}$, showing how any reasonable choice of $N$ cannot limit the coherence of the qubit.

\subsection{Quantum phase slips due to charge fluctuations}

So far we have neglected the impact of the offset charge fluctuations localized on the superconducting islands. It is well known that in the Josephson regime the low energy dynamics of Josephson chains is dominated by quantum phase slips. The latter are sensitive to the random offset charges on the islands via the Aharonov-Casher effect \cite{ivanov2001interference,friedman2002aharonov,matveev2002persistent,pop2012experimental,mizel2020right-sizing}.  The dependence of the spectrum on the charge offsets arises due to the compactness of the phase variables across each Josephson junction. The total Hamiltonian of the system has the general form
\begin{equation}
H=\frac{1}{2}({\bf q}+{\bf q}_{g})^T{\cal C}^{-1}({\bf q}+{\bf q}_{g})+U(\boldsymbol{\varphi}),
\end{equation}
where ${\cal C}$ is the total capacitance matrix of the circuit and  the potential $U$ is the total potential describing the $N_J+1$ junctions, with $N_J=NM$, including also the Blochnium junction of energy $E_J$. The potential $U$ is $2\pi$-periodic in all the phases.  

As discussed in Ref.~\cite{chirolli2006full}, in a superconducting circuit involving only Josephson junctions and capacitances, the compactness of the phases renders the wave function necessarily indistinguishable upon a $2\pi$ phase shift in anyone of the $\varphi_n$ phases, so that the full wave function $\Psi_{\alpha,{\bf q}}$ must be periodic in $\boldsymbol{\varphi}$. This way, the wave function has the general form 
\begin{equation}
\Psi_{l,{\bf q}_g}(\boldsymbol{\varphi})=e^{-i{\bf q}_g\cdot\boldsymbol{\varphi}/2e}\Psi^{\rm Bloch}_{l,{\bf q}_g}(\boldsymbol{\varphi}),
\end{equation}
where $\Psi^{\rm Bloch}_{l,{\bf q}}(\boldsymbol{\varphi})$ is Bloch function. The latter can be easily written in the tight-binding approximation, that is valid when the Josephson energy is much larger than the charging energy, and the resulting Hamiltonian reads
\begin{equation}
H_{l,l'}({\bf q}_g)=h^{(0)}_{l,l'}+\sum_{n}\delta h^{(n)}_{l,l'}2\cos(2\pi q_{g,n}),
\end{equation} 
where $\delta h^{(n)}_{l,l'}$ describes the matrix elements between qubit states mediated by a phase slip in the $n$-th phase $\varphi_n$. By expanding the Hamiltonian in terms of collective modes and employing the resulting eigenstate as a tight-binding basis, we find that a $2\pi$ shift in $n=(j-1)M+k$ phase $\varphi_n$ yields fractional shifts in the collective mode described by ${\bm w}_n$. As a result, the phase-slip-mediated matrix elements can be written as  
\begin{equation}
\delta h^{(n)}_{l,l'}=E^{(0)}_{l'}{\cal O}_{(n)}\int d\varphi\Psi^B_l(\varphi+2\pi)\Psi^B_{l'}(\varphi),
\end{equation}
where the overlap ${\cal O}_{(n)}$ is given by 
\begin{equation}
{\cal O}_{(n)}=\int d\boldsymbol{\gamma}\chi_0(\boldsymbol{\gamma} +{\bm w}_{n}) \chi_0(\boldsymbol{\gamma})=e^{-\frac{\pi^2}{\xi_\gamma^{2}}\frac{N-1}{N}-\frac{\pi^2}{\xi_J^2}\eta_k},
\end{equation}
with $C^{1/2}_k\equiv[W^T(W{\cal C}^{-1}W^T)^{-1/2}W]_{kk}$, $W_{\mu,n}$ the amplitudes of the collective modes, and $\eta_k= C^{1/2}_k/C^{1/2}_J$. By inspection of the capacitance matrix we see that $\eta_k\geq 1$, with $\eta_k\to 1$ for $C_g/C_J\to 0$, so that larger values are obtained for larger ground capacitance $C_g/C_J>1$. As a result we have $e^{-\pi^2\eta_k/\xi_J^2}\leq e^{-\pi^2/\xi_J^2}$, where the right hand side is the quantum phase slip rate of a junction characterized by $\xi_J$. 

The overlap between the Blochnium wave function can be estimated as $e^{-\pi^2\sigma^2}\simeq e^{-\pi^2/N\xi^2_\gamma}$, from which it follows that the quantum phase slip rate of the $k$-th junction in the $j$-th loop is given by
\begin{equation}
\Gamma^{\rm qps}_k=\exp\left[-\frac{\pi^2}{(2M^2)^{1/3}\xi^{4/3}_J}-\frac{\pi^2}{\xi_J^2}\eta_k\right].
\end{equation}
The rate is independent on the number of loops $N$ but it depends on the number of junctions in each loop $M$. By comparing it to the quantum phase slip rate $e^{-\pi^2/\xi_J^2}$, we see that although the effect of the quartic potential results in a looser phase, that can slip out of the minimum with higher probability, the presence of a ground capacitance results in an overall reduced quantum phase slip rate $\Gamma^{\rm qps}_k$. 

The qubit dephasing rate is estimated through Eq.~\eqref{Eq:Gamma-phi} and it is zero at first order at the sweet charge spot ${\bf q}_g=0$. In the worst case scenario in which all charges fluctuate around the value 1/4 (in units of $2e$), by summing the contributions of the independent phase slips the dephasing rate becomes
\begin{equation}
\Gamma_\phi=8\pi AE_{01}N\sum_{k=1}^{M}\Gamma_k^{\rm qps},
\end{equation}
that increases with the number of junctions, in agreement with the result that the superfluid phase in a chain of Josephson junction is destroyed by proliferation of phase slips. In addition, we have that rate for an individual phase slip is slightly increased due to a looser phase in the quartic potential. When expressed by keeping $E_L^{(4)}$ fixed, the dephasing rate becomes
\begin{widetext}
\begin{eqnarray}
\Gamma_\phi&=&8\pi A E_{01}N\sum_{k=1}^M\exp\left\{-\pi^2N\left[\left(\frac{9E_L^{(4)}}{ME_{C_J}}\right)^{1/3}+\eta_k\left(\frac{3MNE_L^{(4)}}{E_{C_J}}\right)^{1/2}\right]\right\}.
\end{eqnarray}
\end{widetext}
It follows that for fixed $E_L^{(4)}$ the dephasing time exponentially increases with $N$ in analogy with the fluxonium \cite{manucharyan2009fluxonium,mizel2020right-sizing}.

\begin{figure}[t]
	\centering
	\includegraphics[width=0.8\linewidth]{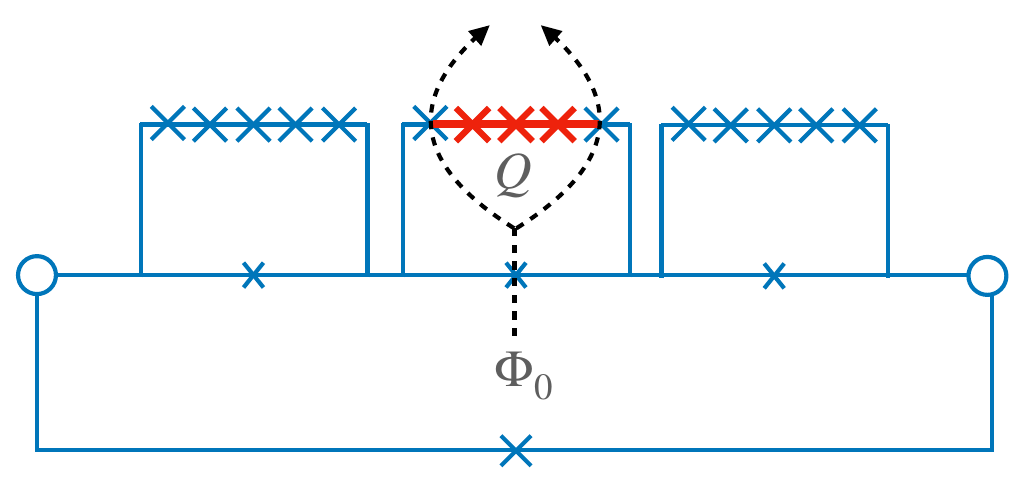}
	\caption{Schematics of the interfereing fluxon tunneling processes that are sensitive to the fluctuating charge on the islands due to the Aharonov-Casher effect. Their amplitude involves tunneling through two junctions, yielding the result Eq.\eqref{Eq:RatioGammaPhi}.}
	\label{Fig7}
\end{figure}

\subsubsection{Comparison with the quadratic Blochnium}

It is instructive to compare the dephasing rate of the quartic Blochnium to the one of the quadratic Blochnium. The dephasing rates can be estimated through a simplified yet powerful argument. The dependence of the spectrum on the offset charges arises due to the Aharonov-Casher effect, in which a fluxon can escape the qubit ring via tunneling through the junctions. Interference between different paths through difference junctions is sensitive to the total charge enclosed between the two junctions. In the quadratic Blochnium, a fluxon can tunnel away from the loop through the chain via one of the $N_J$ large Josephson junctions characterized each by spread $\xi_J$. By independently summing the rates it follows that the quadratic Blochnium has a quantum phase-slip rate given by 
\begin{equation}
\Gamma^{(2)}_{\rm qps}=N_Je^{-\pi^2/\xi_J^2}.
\end{equation}
On the other hand, in the quartic Blochnium a fluxon has first to tunnel via one of the $N$ junctions with Josephson energy $E_{Jc}/M$, that have a quartic potential around each minima and associated spread $\xi_q=\sqrt{2}M^{1/3}\xi_J^{2/3}$, and then through one of the $NM$ junctions with spread $\xi_J$. This is schematically depicted in Fig.~\ref{Fig7}.  Adding the rates we have
\begin{equation}
\Gamma^{(4)}_{\rm qps}=NMe^{-\pi^2/\xi_J^2}e^{-\pi^2/(2M^{2/3}\xi_J^{4/3})}.
\end{equation}
In order to make a comparison meaningful, we assume all energy scales to be the same and only choose $NM=N_J$ in a way to have same qubit frequency $E_{01}$. Choosing the specific values reported in Tab.~\ref{tab:Params}, it follows that the quadratic Blochnium has frequency $E_{01}=\sqrt{E_{Jc}\lambda_0/(2N_J)}\simeq 0.79~{\rm GHz}$, that is of the same order than the quartic Blochnium frequency.  It follows that the ratio of dephasing rates of the qubits is given by
\begin{equation}\label{Eq:RatioGammaPhi}
\frac{\Gamma^{(4)}_\phi}{\Gamma^{(2)}_\phi}=\exp\left[-\frac{\pi^2}{(2M^2)^{1/3}\xi^{4/3}_J}\right],
\end{equation}
where $\Gamma^{(2)}_\phi$ and $\Gamma^{(4)}_\phi$ are the dephasing rates for the quadratic and quartic Blochnium, respectively. It then follows that the quartic one has slightly reduced dephasing rate, because the chain of loops is more robust to charge fluctuations in that the flux has to entirely cross the quarton loops.

\begin{table*}
	\center
	\begin{tabular}{c|c|c|c|c}
	\hline\hline
	$E_J$ (GHz) & $E_C$ (GHz)& $\lambda_0$ (GHz)  & $E_L^{(4)}$ (kHz) & $\omega_{01}$ (GHz)\\
	\hline
	$4.0$ & $ 7.0$ & $3.94$ & $8.13$ & $1.05$ \\
	\hline\hline
 	$E_{Jc}$ (GHz) &  $E_{C_J} $(GHz) & $N,M$ & $\xi/2\pi$  & $\xi_J/2\pi$   \\
	\hline
	$100$ & $ 0.97 $ & $40,8$ & $ 1.64 $ & $0.08$   \\
	\hline\hline
	\end{tabular}
	\caption{Optimal device parameters.}
	\label{tab:Params}
\end{table*}

\section{Conclusions}\label{sect:concl}

In this work we have theoretically discussed a quartic Blochnium, an anharmonic quasi charge superconducting qubit.  The latter is realized through a quartic superinductor closed on a Josephson junction characterized by $E_J\simeq E_C$ in the regime in which the quartic inductance energy scale is $E_L^{(4)}\simeq 10^{-6} E_J,E_C$. The quartic superinductor is formed by a chain of $N$ loops, each constituted by a short chain of $M$ equal Josephson junctions in parallel with a junction whose Josephson energy is $1/M$ times the energy of the other $M$ junctions and by properly threading the loop with half flux quantum $\Phi_0/2$. This single loop realizes the analogous circuit of the quarton qubit \cite{yan2020engineering}, that exploits a quartic Josephson potential to obtain an anharmonic qubit. By considering a short chain of $N$ loops we effectively realize a quartic superinductor described by the potential Eq.~\eqref{Eq:Uphi}. 

The quartic Blochnium we propose shows an anharmonic spectrum, with a ladder of levels similar to the quantum quartic oscillator spectrum, thus solving the weak anharmonicity problem of the Blochnium, that is inherited by the transmon due to its dual character. The quartic anharmonicity is milder than the one achieved in the flux controlled regimes typical of the flux qubit and the fluxonium. Nevertheless, it is less sensitive to flux fluctuations, even accounting for the ones associated to the external flux introduced to achieve the quartic regime. An optimal choice of parameters of the system is provided in Table \ref{tab:Params}. The present design thus promotes the system as a viable and promising quantum computing platform.

{\bf Acknowledgments.} L.C. and F.G. acknowledge EU’s Horizon 2020 Research and Innovation Framework Programme under Grant No. 964398 (SUPERGATE) and No. 101057977 (SPECTRUM).

\appendix

\section{Quantum quartic oscillator spectrum}
\label{App:QuarticO}

The spectrum of a quartic oscillator has no analytic form and we have to resort to numerical diagonalization.  A good strategy is provided by taking the matrix elements of the Hamiltonian Eq.~\eqref{Eq:QuarticH} on the harmonic oscillator states, $\psi_n(\varphi/\xi)$,
\begin{eqnarray}\label{Eq:QHham}
\frac{\langle\psi_n|H|\psi_m\rangle}{(6E_C^2E_L^{(4)})^{1/3}}&=&(3/2+10n/3+n^2)\delta_{nm}\nonumber\\
&+&\frac{n+m+1}{3}(p_m\delta_{n,m+2}+p_n\delta_{n+2,m})\nonumber\\
&+&\frac{1}{6}(p_mp_{m+2}\delta_{n,m+4}+p_np_{n+2}\delta_{n+4,m}),\nonumber\\
\end{eqnarray}
where $p_n=\sqrt{(m+1)(m+2)}$ and diagonalizing the Hamiltonian upon truncation of the spectrum. The nonlinearity of the spectrum is evident from the diagonal terms and the repulsion provided by the off-diagonal terms.

\section{Qubit matrix elements}
\label{App:MatrixEl}

We now need to calculate the matrix elements of the perturbation. The matrix element of the phase $\varphi$ between Bloch states is given by
\begin{eqnarray}
M_{q,s;q',s'}&\equiv&\int_{-\pi}^{\pi} \frac{d\varphi}{2\pi}\psi^*_{q,s}(\varphi)\varphi\psi_{q',s'}(\varphi)\\
&=&i\frac{d}{dq}\delta(q-q')\delta_{ss'}+\delta(q-q')\Omega_{s,s'}(q)\nonumber\\
\Omega_{s,s'}(q)&=&i\int_{-\pi}^\pi\frac{d\varphi}{2\pi}u^*_{qs}(\varphi)\frac{d u_{qs'}(\varphi)}{dq}
\end{eqnarray}
with $\Omega_{s,s}=0$. It follows that for the two lowest energy qubit eigenstates, that belong to the band $s=0$, by  neglecting the contributions coming from $\Omega$ we have that
\begin{eqnarray}
\langle\Psi_{0}|\hat{\varphi}|\Psi_1\rangle &=&i\int_{-1/2}^{1/2}dqv^*_{0}(q)\frac{d v_{1}(q)}{dq},\\
\langle\Psi_{\alpha}|\hat{\varphi}^2|\Psi_\alpha\rangle &=&-\int_{-1/2}^{1/2}dqv^*_{\alpha}(q)\frac{d^2 v_{\alpha}(q)}{dq^2}.
\end{eqnarray}
The integrals can be calculated by constructing approximate Bloch functions $v_l(q)$ in the tight-binding approximation through localized eigenfunctions $\phi_l$. The latter can be chosen as Gaussian eigenstates of the harmonic oscillator, 
\begin{equation}
v_{l,\varphi_x}(q)=\frac{1}{\sqrt{\cal N}}\sum_n e^{-i\varphi_xn}\phi_l(q-n)
\end{equation}
with the spread $\sigma$ to self-consistently determined by minimization of the energy of the Hamiltonian Eq.~\eqref{Eq:DualHam}. The result for the qubit dipole transition matrix element is 
\begin{eqnarray}
\langle\Psi_{0}|\hat{\varphi}|\Psi_1\rangle &=&\frac{i}{\sqrt{2}\sigma}\left[1-\frac{e^{-1/4\sigma^2}}{\sigma^2}\cos(\varphi_x)\right],
\end{eqnarray}
and analogously the other matrix elements.

\section{Chain capacitance matrix}

The capacitance matrix $C$ of the quarton is an $(M+1)\times(M+1)$ real, symmetric, positive-definite matrix, with entries $C_{n,n}=C_g+2C_{Jc}$ for $n=1,\ldots,M-1$, and $C_{n,n+1}=C_{n+1,n}=-C_{Jc}$ for $n=0,\ldots, M-1$, $C_{0,0}=C_{M,M}=C_g+(1+\alpha)C_{Jc}$, $C_{0,M}=C_{M,0}=-\alpha C_{Jc}$. Here, $C_{Jc}$ ($\alpha C_{Jc}$) is the capacitance of each junction with energy $E_{Jc}$ ($\alpha E_{Jc}$), and $C_g$ is the capacitance of each island with the ground reference \cite{vool2017introduction,krantz2019quantum}. The capacitance matrix of the independent $M$ degrees of freedom is given by $[{\cal C}^{-1}]_{n,m}=(JC^{-1}J^T)_{n,m}$ with $n,m=1,\ldots,M$, where the matrix $J$ specified by the entries $J_{n,m}=-\delta_{n,m}+\delta_{n+1,m}$ for $1\leq n<M+1$ and $J_{M+1,M+1}=1$. The capacitance matrix of the chain of quarton is similarly obtained.

\bibliographystyle{quantum}
\bibliography{biblioQSI}

\newcommand{\PageBackRef}{}
\PageBackRef
\begin{thebibliography}{10}

\bibitem{kjaergaard2020superconducting}
Morten Kjaergaard, Mollie~E. Schwartz, Jochen Braum\"{u}ller, Philip Krantz,
  Joel I.-J. Wang, Simon Gustavsson, and William~D. Oliver.
\newblock ``Superconducting qubits: Current state of play''.
\newblock
  \href{https://doi.org/10.1146/annurev-conmatphys-031119-050605}{Annual Review
  of Condensed Matter Physics {\bf 11}, 369--395}~(2020). Appearances:~

\bibitem{huang2020superconducting}
He-Liang Huang, Dachao Wu, Daojin Fan, and Xiaobo Zhu.
\newblock ``Superconducting quantum computing: a review''.
\newblock \href{https://doi.org/10.1007/s11432-020-2881-9}{Science China
  Information Sciences {\bf 63}, 180501}~(2020). Appearances:~

\bibitem{siddiqi2021engineering}
Irfan Siddiqi.
\newblock ``Engineering high-coherence superconducting qubits''.
\newblock \href{https://doi.org/10.1038/s41578-021-00370-4}{Nature Reviews
  Materials {\bf 6}, 875--891}~(2021). Appearances:~

\bibitem{bravyi2022future}
Sergey Bravyi, Oliver Dial, Jay~M. Gambetta, Dar{\'\i}o Gil, and Zaira Nazario.
\newblock ``The future of quantum computing with superconducting qubits''.
\newblock \href{https://doi.org/10.1063/5.0082975}{Journal of Applied Physics
  {\bf 132}, 160902}~(2022). Appearances:~

\bibitem{gottesman1997stabilizer}
Daniel {Gottesman}.
\newblock ``{Stabilizer codes and quantum error correction}''.
\newblock \href{https://doi.org/10.7907/rzr7-dt72}{PhD thesis}.
\newblock California Institute of Technology.
\newblock ~(1997). Appearances:~

\bibitem{nakamura1999coherent}
Y.~Nakamura, Yu.~A. Pashkin, and J.~S. Tsai.
\newblock ``Coherent control of macroscopic quantum states in a
  single-cooper-pair box''.
\newblock \href{https://doi.org/10.1038/19718}{Nature {\bf 398},
  786--788}~(1999). Appearances:~

\bibitem{gambetta2017building}
Jay~M. Gambetta, Jerry~M. Chow, and Matthias Steffen.
\newblock ``Building logical qubits in a superconducting quantum computing
  system''.
\newblock \href{https://doi.org/10.1038/s41534-016-0004-0}{npj Quantum
  Information {\bf 3}, 2}~(2017). Appearances:~

\bibitem{koch2007charge-insensitive}
Jens Koch, Terri~M. Yu, Jay Gambetta, A.~A. Houck, D.~I. Schuster, J.~Majer,
  Alexandre Blais, M.~H. Devoret, S.~M. Girvin, and R.~J. Schoelkopf.
\newblock ``Charge-insensitive qubit design derived from the cooper pair box''.
\newblock \href{https://doi.org/10.1103/PhysRevA.76.042319}{Phys. Rev. A {\bf
  76}, 042319}~(2007). Appearances:~

\bibitem{manucharyan2009fluxonium}
Vladimir~E. Manucharyan, Jens Koch, Leonid~I. Glazman, and Michel~H. Devoret.
\newblock ``Fluxonium: Single cooper-pair circuit free of charge offsets''.
\newblock \href{https://doi.org/10.1126/science.1175552}{Science {\bf 326},
  113--116}~(2009). Appearances:~

\bibitem{lin2018demonstration}
Yen-Hsiang Lin, Long~B. Nguyen, Nicholas Grabon, Jonathan San~Miguel, Natalia
  Pankratova, and Vladimir~E. Manucharyan.
\newblock ``Demonstration of protection of a superconducting qubit from energy
  decay''.
\newblock \href{https://doi.org/10.1103/PhysRevLett.120.150503}{Phys. Rev.
  Lett. {\bf 120}, 150503}~(2018). Appearances:~

\bibitem{nguyenPRX2019high-coherence}
Long~B. Nguyen, Yen-Hsiang Lin, Aaron Somoroff, Raymond Mencia, Nicholas
  Grabon, and Vladimir~E. Manucharyan.
\newblock ``High-coherence fluxonium qubit''.
\newblock \href{https://doi.org/10.1103/PhysRevX.9.041041}{Phys. Rev. X {\bf
  9}, 041041}~(2019). Appearances:~

\bibitem{somoroff2021millisecond}
Aaron Somoroff, Quentin Ficheux, Raymond~A. Mencia, Haonan Xiong, Roman Kuzmin,
  and Vladimir~E. Manucharyan.
\newblock ``Millisecond coherence in a superconducting qubit''.
\newblock \href{https://doi.org/10.1103/PhysRevLett.130.267001}{Phys. Rev.
  Lett. {\bf 130}, 267001}~(2023). Appearances:~

\bibitem{zhang2021universal}
Helin Zhang, Srivatsan Chakram, Tanay Roy, Nathan Earnest, Yao Lu, Ziwen Huang,
  D.~K. Weiss, Jens Koch, and David~I. Schuster.
\newblock ``Universal fast-flux control of a coherent, low-frequency qubit''.
\newblock \href{https://doi.org/10.1103/PhysRevX.11.011010}{Phys. Rev. X {\bf
  11}, 011010}~(2021). Appearances:~

\bibitem{nguyen2022blueprint}
Long~B. Nguyen, Gerwin Koolstra, Yosep Kim, Alexis Morvan, Trevor Chistolini,
  Shraddha Singh, Konstantin~N. Nesterov, Christian J\"unger, Larry Chen, Zahra
  Pedramrazi, Bradley~K. Mitchell, John~Mark Kreikebaum, Shruti Puri, David~I.
  Santiago, and Irfan Siddiqi.
\newblock ``Blueprint for a high-performance fluxonium quantum processor''.
\newblock \href{https://doi.org/10.1103/PRXQuantum.3.037001}{PRX Quantum {\bf
  3}, 037001}~(2022). Appearances:~

\bibitem{orlando1999superconducting}
T.~P. Orlando, J.~E. Mooij, Lin Tian, Caspar~H. van~der Wal, L.~S. Levitov,
  Seth Lloyd, and J.~J. Mazo.
\newblock ``Superconducting persistent-current qubit''.
\newblock \href{https://doi.org/10.1103/PhysRevB.60.15398}{Phys. Rev. B {\bf
  60}, 15398--15413}~(1999). Appearances:~

\bibitem{manucharyan2012thesis}
V.~E. Manucharyan.
\newblock ``Superinductance''.
\newblock PhD thesis.
\newblock Yale University.
\newblock ~(2012). Appearances:~

\bibitem{lloyd1999quantum}
Seth Lloyd and Samuel~L. Braunstein.
\newblock ``Quantum computation over continuous variables''.
\newblock \href{https://doi.org/10.1103/PhysRevLett.82.1784}{Phys. Rev. Lett.
  {\bf 82}, 1784--1787}~(1999). Appearances:~

\bibitem{gottesman2001encoding}
Daniel Gottesman, Alexei Kitaev, and John Preskill.
\newblock ``Encoding a qubit in an oscillator''.
\newblock \href{https://doi.org/10.1103/PhysRevA.64.012310}{Phys. Rev. A {\bf
  64}, 012310}~(2001). Appearances:~

\bibitem{ofek2016extending}
Nissim Ofek, Andrei Petrenko, Reinier Heeres, Philip Reinhold, Zaki Leghtas,
  Brian Vlastakis, Yehan Liu, Luigi Frunzio, S.~M. Girvin, L.~Jiang, Mazyar
  Mirrahimi, M.~H. Devoret, and R.~J. Schoelkopf.
\newblock ``Extending the lifetime of a quantum bit with error correction in
  superconducting circuits''.
\newblock \href{https://doi.org/10.1038/nature18949}{Nature {\bf 536},
  441--445}~(2016). Appearances:~

\bibitem{hu2019quantum}
L.~Hu, Y.~Ma, W.~Cai, X.~Mu, Y.~Xu, W.~Wang, Y.~Wu, H.~Wang, Y.~P. Song, C.~L.
  Zou, S.~M. Girvin, L-M. Duan, and L.~Sun.
\newblock ``Quantum error correction and universal gate set operation on a
  binomial bosonic logical qubit''.
\newblock \href{https://doi.org/10.1038/s41567-018-0414-3}{Nature Physics {\bf
  15}, 503--508}~(2019). Appearances:~

\bibitem{campagne-Ibarcq2020quantum}
P.~Campagne-Ibarcq, A.~Eickbusch, S.~Touzard, E.~Zalys-Geller, N.~E. Frattini,
  V.~V. Sivak, P.~Reinhold, S.~Puri, S.~Shankar, R.~J. Schoelkopf, L.~Frunzio,
  M.~Mirrahimi, and M.~H. Devoret.
\newblock ``Quantum error correction of a qubit encoded in grid states of an
  oscillator''.
\newblock \href{https://doi.org/10.1038/s41586-020-2603-3}{Nature {\bf 584},
  368--372}~(2020). Appearances:~

\bibitem{blatter2001design}
Gianni Blatter, Vadim~B. Geshkenbein, and Lev~B. Ioffe.
\newblock ``Design aspects of superconducting-phase quantum bits''.
\newblock \href{https://doi.org/10.1103/PhysRevB.63.174511}{Phys. Rev. B {\bf
  63}, 174511}~(2001). Appearances:~

\bibitem{protopopov2004anomalous}
Ivan~V. Protopopov and Mikhail~V. Feigel'man.
\newblock ``Anomalous periodicity of supercurrent in long frustrated
  josephson-junction rhombi chains''.
\newblock \href{https://doi.org/10.1103/PhysRevB.70.184519}{Phys. Rev. B {\bf
  70}, 184519}~(2004). Appearances:~

\bibitem{kitaev2006protected}
Alexei {Kitaev}.
\newblock ``{Protected qubit based on a superconducting current
  mirror}''~(2006).
\newblock
  \href{https://doi.org/10.48550/arXiv.cond-mat/0609441}{arXiv:cond-mat/0609441}.
  Appearances:~

\bibitem{gladchenko2009superconducting}
Sergey Gladchenko, David Olaya, Eva Dupont-Ferrier, Benoit Dou{\c c}ot, Lev~B.
  Ioffe, and Michael~E. Gershenson.
\newblock ``Superconducting nanocircuits for topologically protected qubits''.
\newblock \href{https://doi.org/10.1038/nphys1151}{Nature Physics {\bf 5},
  48--53}~(2009). Appearances:~

\bibitem{smith2020superconducting}
W.~C. Smith, A.~Kou, X.~Xiao, U.~Vool, and M.~H. Devoret.
\newblock ``Superconducting circuit protected by two-cooper-pair tunneling''.
\newblock \href{https://doi.org/10.1038/s41534-019-0231-2}{npj Quantum
  Information {\bf 6}, 8}~(2020). Appearances:~

\bibitem{gyenis2021experimental}
Andr\'as Gyenis, Pranav~S. Mundada, Agustin Di~Paolo, Thomas~M. Hazard, Xinyuan
  You, David~I. Schuster, Jens Koch, Alexandre Blais, and Andrew~A. Houck.
\newblock ``Experimental realization of a protected superconducting circuit
  derived from the $0$--$\ensuremath{\pi}$ qubit''.
\newblock \href{https://doi.org/10.1103/PRXQuantum.2.010339}{PRX Quantum {\bf
  2}, 010339}~(2021). Appearances:~

\bibitem{chirolli2021enhanced}
Luca Chirolli and Joel~E. Moore.
\newblock ``Enhanced coherence in superconducting circuits via band
  engineering''.
\newblock \href{https://doi.org/10.1103/PhysRevLett.126.187701}{Phys. Rev.
  Lett. {\bf 126}, 187701}~(2021). Appearances:~

\bibitem{chirolli2022swap}
Luca Chirolli, Norman~Y. Yao, and Joel~E. Moore.
\newblock ``Swap gate between a majorana qubit and a parity-protected
  superconducting qubit''.
\newblock \href{https://doi.org/10.1103/PhysRevLett.129.177701}{Phys. Rev.
  Lett. {\bf 129}, 177701}~(2022). Appearances:~

\bibitem{smith2022magnifying}
W.~C. Smith, M.~Villiers, A.~Marquet, J.~Palomo, M.~R. Delbecq, T.~Kontos,
  P.~Campagne-Ibarcq, B.~Dou\ifmmode~\mbox{\c{c}}\else \c{c}\fi{}ot, and
  Z.~Leghtas.
\newblock ``Magnifying quantum phase fluctuations with cooper-pair pairing''.
\newblock \href{https://doi.org/10.1103/PhysRevX.12.021002}{Phys. Rev. X {\bf
  12}, 021002}~(2022). Appearances:~

\bibitem{calzona2023multi-mode}
Alessio Calzona and Matteo Carrega.
\newblock ``Multi-mode architectures for noise-resilient superconducting
  qubits''.
\newblock \href{https://doi.org/10.1088/1361-6668/acaa64}{Superconductor
  Science and Technology {\bf 36}, 023001}~(2022). Appearances:~

\bibitem{calzona2023anomalous}
Alessio Calzona, Matteo Carrega, and Luca Chirolli.
\newblock ``Anomalous periodicity and parafermion hybridization in
  superconducting qubits''.
\newblock \href{https://doi.org/10.1103/PhysRevB.107.045105}{Phys. Rev. B {\bf
  107}, 045105}~(2023). Appearances:~

\bibitem{kalashnikov2020bifluxon}
Konstantin Kalashnikov, Wen~Ting Hsieh, Wenyuan Zhang, Wen-Sen Lu, Plamen
  Kamenov, Agustin Di~Paolo, Alexandre Blais, Michael~E. Gershenson, and
  Matthew Bell.
\newblock ``Bifluxon: Fluxon-parity-protected superconducting qubit''.
\newblock \href{https://doi.org/10.1103/PRXQuantum.1.010307}{PRX Quantum {\bf
  1}, 010307}~(2020). Appearances:~

\bibitem{bell2012quantum}
M.~T. Bell, I.~A. Sadovskyy, L.~B. Ioffe, A.~Yu. Kitaev, and M.~E. Gershenson.
\newblock ``Quantum superinductor with tunable nonlinearity''.
\newblock \href{https://doi.org/10.1103/PhysRevLett.109.137003}{Phys. Rev.
  Lett. {\bf 109}, 137003}~(2012). Appearances:~

\bibitem{bell2018josephson}
Matthew~T. Bell, Beno{\^\i}t Dou{\c c}ot, Michael~E. Gershenson, Lev~B. Ioffe,
  and Aleksandra Petkovi{\'c}.
\newblock ``Josephson ladders as a model system for 1d quantum phase
  transitions''.
\newblock
  \href{https://doi.org/https://doi.org/10.1016/j.crhy.2018.09.002}{Comptes
  Rendus Physique {\bf 19}, 484--497}~(2018). Appearances:~

\bibitem{peruzzo2020surpassing}
M.~Peruzzo, A.~Trioni, F.~Hassani, M.~Zemlicka, and J.~M. Fink.
\newblock ``Surpassing the resistance quantum with a geometric superinductor''.
\newblock \href{https://doi.org/10.1103/PhysRevApplied.14.044055}{Phys. Rev.
  Appl. {\bf 14}, 044055}~(2020). Appearances:~

\bibitem{feng-ming2021quantum}
Feng-Ming {Liu}, Ming-Cheng {Chen}, Can {Wang}, Shao-Wei {Li}, Zhong-Xia
  {Shang}, Chong {Ying}, Jian-Wen {Wang}, Cheng-Zhi {Peng}, Xiaobo {Zhu},
  Chao-Yang {Lu}, and Jian-Wei {Pan}.
\newblock ``{Quantum design for advanced qubits: plasmonium}''~(2021).
\newblock \href{https://doi.org/10.48550/arXiv.2109.00994}{arXiv:2109.00994}.
  Appearances:~

\bibitem{peruzzo2021geometric}
Matilda Peruzzo, Farid Hassani, Gregory Szep, Andrea Trioni, Elena Redchenko,
  Martin \ifmmode \check{Z}\else \v{Z}\fi{}emli\ifmmode~\check{c}\else
  \v{c}\fi{}ka, and Johannes~M. Fink.
\newblock ``Geometric superinductance qubits: Controlling phase delocalization
  across a single josephson junction''.
\newblock \href{https://doi.org/10.1103/PRXQuantum.2.040341}{PRX Quantum {\bf
  2}, 040341}~(2021). Appearances:~

\bibitem{ranadive2022kerr}
Arpit Ranadive, Martina Esposito, Luca Planat, Edgar Bonet, C{\'e}cile Naud,
  Olivier Buisson, Wiebke Guichard, and Nicolas Roch.
\newblock ``Kerr reversal in josephson meta-material and traveling wave
  parametric amplification''.
\newblock \href{https://doi.org/10.1038/s41467-022-29375-5}{Nature
  Communications {\bf 13}, 1737}~(2022). Appearances:~

\bibitem{koch2009charging}
Jens Koch, V.~Manucharyan, M.~H. Devoret, and L.~I. Glazman.
\newblock ``Charging effects in the inductively shunted josephson junction''.
\newblock \href{https://doi.org/10.1103/PhysRevLett.103.217004}{Phys. Rev.
  Lett. {\bf 103}, 217004}~(2009). Appearances:~

\bibitem{pechenezhskiy2020superconducting}
Ivan~V. Pechenezhskiy, Raymond~A. Mencia, Long~B. Nguyen, Yen-Hsiang Lin, and
  Vladimir~E. Manucharyan.
\newblock ``The superconducting quasicharge qubit''.
\newblock \href{https://doi.org/10.1038/s41586-020-2687-9}{Nature {\bf 585},
  368--371}~(2020). Appearances:~

\bibitem{herrera-marti2013tradeoff}
David~A. Herrera-Mart\'{\i}, Ahsan Nazir, and Sean~D. Barrett.
\newblock ``Tradeoff between leakage and dephasing errors in the fluxonium
  qubit''.
\newblock \href{https://doi.org/10.1103/PhysRevB.88.094512}{Phys. Rev. B {\bf
  88}, 094512}~(2013). Appearances:~

\bibitem{zorin2009superconducting}
A.~B. Zorin and F.~Chiarello.
\newblock ``Superconducting phase qubit based on the josephson oscillator with
  strong anharmonicity''.
\newblock \href{https://doi.org/10.1103/PhysRevB.80.214535}{Phys. Rev. B {\bf
  80}, 214535}~(2009). Appearances:~

\bibitem{yan2020engineering}
Fei {Yan}, Youngkyu {Sung}, Philip {Krantz}, Archana {Kamal}, David~K. {Kim},
  Jonilyn~L. {Yoder}, Terry~P. {Orlando}, Simon {Gustavsson}, and William~D.
  {Oliver}.
\newblock ``{Engineering Framework for Optimizing Superconducting Qubit
  Designs}''~(2020).
\newblock \href{https://doi.org/10.48550/arXiv.2006.04130}{arXiv:2006.04130}.
  Appearances:~

\bibitem{hyyppa2022unimon}
Eric Hyypp{\"a}, Suman Kundu, Chun~Fai Chan, Andr{\'a}s Gunyh{\'o}, Juho
  Hotari, David Janzso, Kristinn Juliusson, Olavi Kiuru, Janne Kotilahti,
  Alessandro Landra, Wei Liu, Fabian Marxer, Akseli M{\"a}kinen, Jean-Luc
  Orgiazzi, Mario Palma, Mykhailo Savytskyi, Francesca Tosto, Jani Tuorila,
  Vasilii Vadimov, Tianyi Li, Caspar Ockeloen-Korppi, Johannes Heinsoo,
  Kuan~Yen Tan, Juha Hassel, and Mikko M{\"o}tt{\"o}nen.
\newblock ``Unimon qubit''.
\newblock \href{https://doi.org/10.1038/s41467-022-34614-w}{Nature
  Communications {\bf 13}, 6895}~(2022). Appearances:~

\bibitem{matveev2002persistent}
K.~A. Matveev, A.~I. Larkin, and L.~I. Glazman.
\newblock ``Persistent current in superconducting nanorings''.
\newblock \href{https://doi.org/10.1103/PhysRevLett.89.096802}{Phys. Rev. Lett.
  {\bf 89}, 096802}~(2002). Appearances:~

\bibitem{brooks2013protected}
Peter Brooks, Alexei Kitaev, and John Preskill.
\newblock ``Protected gates for superconducting qubits''.
\newblock \href{https://doi.org/10.1103/PhysRevA.87.052306}{Phys. Rev. A {\bf
  87}, 052306}~(2013). Appearances:~

\bibitem{masluk2012microwave}
Nicholas~A. Masluk, Ioan~M. Pop, Archana Kamal, Zlatko~K. Minev, and Michel~H.
  Devoret.
\newblock ``Microwave characterization of josephson junction arrays:
  Implementing a low loss superinductance''.
\newblock \href{https://doi.org/10.1103/PhysRevLett.109.137002}{Phys. Rev.
  Lett. {\bf 109}, 137002}~(2012). Appearances:~

\bibitem{niepce2019high}
David Niepce, Jonathan Burnett, and Jonas Bylander.
\newblock ``High kinetic inductance $\mathrm{Nb}\mathrm{N}$ nanowire
  superinductors''.
\newblock \href{https://doi.org/10.1103/PhysRevApplied.11.044014}{Phys. Rev.
  Appl. {\bf 11}, 044014}~(2019). Appearances:~

\bibitem{kuzmin2019quantum}
R.~Kuzmin, R.~Mencia, N.~Grabon, N.~Mehta, Y.~H. Lin, and V.~E. Manucharyan.
\newblock ``Quantum electrodynamics of a superconductor--insulator phase
  transition''.
\newblock \href{https://doi.org/10.1038/s41567-019-0553-1}{Nature Physics {\bf
  15}, 930--934}~(2019). Appearances:~

\bibitem{grunhaupt2019granular}
Lukas Gr{\"u}nhaupt, Martin Spiecker, Daria Gusenkova, Nataliya Maleeva,
  Sebastian~T. Skacel, Ivan Takmakov, Francesco Valenti, Patrick Winkel, Hannes
  Rotzinger, Wolfgang Wernsdorfer, Alexey~V. Ustinov, and Ioan~M. Pop.
\newblock ``Granular aluminium as a superconducting material for high-impedance
  quantum circuits''.
\newblock \href{https://doi.org/10.1038/s41563-019-0350-3}{Nature Materials
  {\bf 18}, 816--819}~(2019). Appearances:~

\bibitem{kamenov2020granular}
Plamen Kamenov, Wen-Sen Lu, Konstantin Kalashnikov, Thomas DiNapoli, Matthew~T.
  Bell, and Michael~E. Gershenson.
\newblock ``Granular aluminum meandered superinductors for quantum circuits''.
\newblock \href{https://doi.org/10.1103/PhysRevApplied.13.054051}{Phys. Rev.
  Appl. {\bf 13}, 054051}~(2020). Appearances:~

\bibitem{fergusonPRX2013symmetries}
David~G. Ferguson, A.~A. Houck, and Jens Koch.
\newblock ``Symmetries and collective excitations in large superconducting
  circuits''.
\newblock \href{https://doi.org/10.1103/PhysRevX.3.011003}{Phys. Rev. X {\bf
  3}, 011003}~(2013). Appearances:~

\bibitem{weiss2015kerr}
T.~Wei\ss{}l, B.~K\"ung, E.~Dumur, A.~K. Feofanov, I.~Matei, C.~Naud,
  O.~Buisson, F.~W.~J. Hekking, and W.~Guichard.
\newblock ``Kerr coefficients of plasma resonances in josephson junction
  chains''.
\newblock \href{https://doi.org/10.1103/PhysRevB.92.104508}{Phys. Rev. B {\bf
  92}, 104508}~(2015). Appearances:~

\bibitem{chirolli2006decoherence}
Luca Chirolli and Guido Burkard.
\newblock ``Decoherence in solid-state qubits''.
\newblock \href{https://doi.org/10.1080/00018730802218067}{Advances in Physics
  {\bf 57}, 225--285}~(2008). Appearances:~

\bibitem{yan2016theflux}
Fei Yan, Simon Gustavsson, Archana Kamal, Jeffrey Birenbaum, Adam~P Sears,
  David Hover, Ted~J. Gudmundsen, Danna Rosenberg, Gabriel Samach, S.~Weber,
  Jonilyn~L. Yoder, Terry~P. Orlando, John Clarke, Andrew~J. Kerman, and
  William~D. Oliver.
\newblock ``The flux qubit revisited to enhance coherence and
  reproducibility''.
\newblock \href{https://doi.org/10.1038/ncomms12964}{Nature Communications {\bf
  7}, 12964}~(2016). Appearances:~

\bibitem{ivanov2001interference}
Dmitri~A. Ivanov, Lev~B. Ioffe, Vadim~B. Geshkenbein, and Gianni Blatter.
\newblock ``Interference effects in isolated josephson junction arrays with
  geometric symmetries''.
\newblock \href{https://doi.org/10.1103/PhysRevB.65.024509}{Phys. Rev. B {\bf
  65}, 024509}~(2001). Appearances:~

\bibitem{friedman2002aharonov}
Jonathan~R. Friedman and D.~V. Averin.
\newblock ``Aharonov-casher-effect suppression of macroscopic tunneling of
  magnetic flux''.
\newblock \href{https://doi.org/10.1103/PhysRevLett.88.050403}{Phys. Rev. Lett.
  {\bf 88}, 050403}~(2002). Appearances:~

\bibitem{pop2012experimental}
I.~M. Pop, B.~Dou\ifmmode~\mbox{\c{c}}\else \c{c}\fi{}ot, L.~Ioffe,
  I.~Protopopov, F.~Lecocq, I.~Matei, O.~Buisson, and W.~Guichard.
\newblock ``Experimental demonstration of aharonov-casher interference in a
  josephson junction circuit''.
\newblock \href{https://doi.org/10.1103/PhysRevB.85.094503}{Phys. Rev. B {\bf
  85}, 094503}~(2012). Appearances:~

\bibitem{mizel2020right-sizing}
Ari Mizel and Yariv Yanay.
\newblock ``Right-sizing fluxonium against charge noise''.
\newblock \href{https://doi.org/10.1103/PhysRevB.102.014512}{Phys. Rev. B {\bf
  102}, 014512}~(2020). Appearances:~

\bibitem{chirolli2006full}
Luca Chirolli and Guido Burkard.
\newblock ``Full control of qubit rotations in a voltage-biased superconducting
  flux qubit''.
\newblock \href{https://doi.org/10.1103/PhysRevB.74.174510}{Phys. Rev. B {\bf
  74}, 174510}~(2006). Appearances:~

\bibitem{vool2017introduction}
Uri Vool and Michel Devoret.
\newblock ``Introduction to quantum electromagnetic circuits''.
\newblock \href{https://doi.org/https://doi.org/10.1002/cta.2359}{International
  Journal of Circuit Theory and Applications {\bf 45}, 897--934}~(2017).
  Appearances:~

\bibitem{krantz2019quantum}
P.~Krantz, M.~Kjaergaard, F.~Yan, T.~P. Orlando, S.~Gustavsson, and W.~D.
  Oliver.
\newblock ``A quantum engineer's guide to superconducting qubits''.
\newblock \href{https://doi.org/10.1063/1.5089550}{Applied Physics Reviews {\bf
  6}, 021318}~(2019). Appearances:~

\end{thebibliography}

\end{document}